\documentclass[aps,prl,twocolumn,showpacs,groupedaddress,floatfix]{revtex4}
\usepackage{amsmath}
\usepackage{graphicx}

\begin{document}

\newcommand{\jimhsection}[1]{\vspace{3pt} \noindent {\bf{#1}.}}

\title{Enhancing practical security of quantum key distribution with a few decoy states}

\author{Jim W. Harrington}
\email[]{jimh@lanl.gov}
\altaffiliation{Mail Stop D454}
\author{J. Mark Ettinger}
\email[]{ettinger@lanl.gov}
\altaffiliation{Mail Stop B230}
\author{Richard J. Hughes}
\email[]{hughes@lanl.gov}
\altaffiliation{Mail Stop D454}
\author{Jane E. Nordholt}
\email[]{jnordholt@lanl.gov}
\altaffiliation{Mail Stop D454}
\affiliation{Los Alamos National Laboratory, Los Alamos, NM 87545}

\date{\today}

\begin{abstract}
Quantum key distribution establishes a secret string of bits between two distant parties. Of concern in weak laser pulse schemes is the especially strong photon number splitting attack by an eavesdropper, but the decoy state method can detect this attack with current technology, yielding a high rate of secret bits. In this Letter, we develop rigorous security statements in the case of finite statistics with only a few decoy states, and we present the results of simulations of an experimental setup of a decoy state protocol that can be simply realized with current technology.
\end{abstract}


\maketitle

\jimhsection{Introduction}
Tasks can be performed with quantum information processing that are difficult or impossible by purely classical means.  Quantum key distribution (QKD) establishes secret keys shared between separated parties to enable secure communication, by making use of a quantum channel and a public authenticated classical channel.  In the BB84 QKD protocol \cite{bb84} random bits are encoded into polarized single-photon signals sent between the two parties (traditionally named Alice and Bob); a rigorous upper bound on the information gain of any potential eavesdropper (Eve) is deduced by measuring the bit error rate (BER) of the quantum signals.  This information is then erased by privacy amplification \cite{pa_robert,pa_bennett} via public communications between Alice and Bob over the classical channel with universal hashing \cite{univ_hash_1, univ_hash_2}, producing a shared secret cryptographic key. However, in practice, QKD is popularly implemented with highly-attenuated weak laser pulse quantum signals, which are characterized by a Poissonian photon number probability distribution with mean $\mu < 1$.  Thus, with probability $1 - e^{-\mu} (1+\mu) \sim O(\mu^2/2)$, Alice prepares a pulse containing more than one photon. Furthermore, there is typically considerable loss in the Alice-Bob quantum channel, amounting to a 10-20 dB attenuation in many experiments. An eavesdropper could hypothetically exploit this loss, in conjunction with the multi-photon signals \cite{weak_coherent_qkd}, with very strong attacks such as photon number splitting (PNS) \cite{limitations_qkd}, to gain information on the final key, unless $\mu$ is chosen to be sufficiently small. However, the recent invention of the decoy state method \cite{hwang_decoy, lo_decoy} provides a rigorous means to foil this class of attacks with $\mu \sim O(1)$, elevating the security of weak laser pulse QKD.

In a PNS attack Eve performs a photon number non-demolition measurement to identify Alice's multi-photon signals.  Eve removes one photon from each multi-photon signal and stores it in a quantum memory, while sending on the signal's remaining photons to Bob over a lower-loss quantum channel. Eve also blocks enough of Alice's single-photon signals so that Bob does not notice any change in bit rate. Then, when the measurement bases are revealed during the sifting stage of the BB84 protocol Eve could obtain complete knowledge of the stored photons without introducing any statistical disturbance in the Alice-Bob quantum communications \cite{realistic_qkd}. With sufficient loss in the Alice-Bob quantum channel, Eve could block all of Alice's single-photon signals and learn the entire key. However, decoy state protocols allow Alice and Bob to thwart the hypothetical PNS attack, and other attacks exploiting channel loss and multi-photon signals, on weak laser pulse QKD by enabling them to establish a lower bound on the fraction of bits in the sifted key that originated from Alice as single-photon signals. Privacy amplification may then be used to obtain a secret key, making the conservative assumption that all multi-photon signals are known to Eve.

In a decoy state protocol, Alice randomly selects the mean photon number of each of her pulses from among a set of values between $\mu_{low}$ and $\mu_{high}$.  Each pulse encodes a random bit in a random basis of an orthogonal space (such as polarization or phase).  Alice and Bob (publicly) count the number of detection events (clicks signifying that one or more photons were received) for each level.  If ``too many'' detections occur at the high levels and ``too few'' at the low levels, Alice and Bob may suspect they are victims of a photon number splitting attack.  This is made rigorous and shown to be asymptotically efficient in \cite{lo_decoy}.  Some approaches towards developing a practical protocol are presented in \cite{wang_decoy, lo_practical}.

The main result of this paper is a security statement and protocol for the 
decoy state method for the case of finite samples; it could be incorporated 
into an experiment to get a real-world security measure for a finite-length 
final secret key.  Our approach \cite{lanl_protocol} is to develop 
security statements of the form, ``With confidence $1-\epsilon$, Eve's 
distribution over final $m$-bit keys has Shannon entropy at least $m-1$.''  
That is, the a priori probability that Eve has less than $m-1$ bits of 
Shannon entropy for the final key shared by Alice and Bob is less than $\epsilon$.  However, 
the security statement we present here will be limited to the number of 
bits Alice and Bob share from single-photon pulses.  A complete security 
statement could then be constructed by integrating the steps of 
error reconciliation, privacy amplification, authentication, and key 
verification with appropriate confidence levels.

\jimhsection{Analysis}
A signal from Alice with mean photon number $\mu$ is detected by Bob 
with probability $d_{\mu} = \sum_{n \geq 0} \frac{e^{-\mu} \mu^n}{n!} y_n$,
where the unknowns $\{y_n\}$ represent the channel transmission 
characteristics, with $0 \leq y_n \leq 1$.  More precisely, $y_n$ is the 
conditional probability that at least one photon is detected given that $n$ 
photons were emitted.  Now suppose that Alice utilizes $M$ mean photon 
numbers, $\mu_1,\mu_2,...,\mu_M.$  For each $\mu_j$, Alice's detection 
data (from a beam monitor) provides not only a maximum likelihood 
estimator $\hat{\mu_j}$, but also, more importantly for our purposes, a
$1-\epsilon$ confidence interval 
\begin{eqnarray}
X^-_j \leq \mu_j \leq X^+_j\,.
\end{eqnarray}
Similarly, for each $\mu_j$, Bob's detection data yields a $1 - \epsilon$
confidence interval for $d_{\mu_j}$.  By truncating the infinite
series of $d_{\mu_j}$ after ($K$+1) terms, with bounds for the dropped portion, 
we obtain $2M$ inequalities of the form
\begin{eqnarray}
Y^-_j \leq \sum_{0 \leq k \leq K} \frac{e^{-\mu_j} (\mu_j)^k}{k!} y_k \leq Y^+_j\,.
\end{eqnarray}
We choose $K$ sufficiently large to achieve tight bounds (limited only by computational power).
We want to conservatively bound the unknowns $\{\mu_j\}_{j \leq M}$ and
$\{y_k \}_{k \leq K}$ utilizing these $4M$ inequalities and the $2(K + 1)$ trivial 
inequalities $0 \leq y_k \leq 1$.  Let the closed, bounded region in the 
$(M + K + 1)$-dimensional real vector space defined by all $4M + 2(K + 1)$ 
inequalities be denoted $R$.
Note that the parameters of interest $\{\mu_j\}_{j \leq M}$,$\{y_k
\}_{k \leq K}$ lie in $R$ with confidence $(1 - \epsilon)^ {2M}.$
The conditional probability of a {\em single} photon detection
conditioned on a detection at Bob for a fixed mean photon number
is
\begin{eqnarray}
P(\mu,y_0,y_1,...) 
&=& \frac{e^{-\mu} \mu y_1}{\sum_{n \geq 0} \frac{e^{-\mu} \mu^n}{n!} y_n}\,.
\end{eqnarray} 

Let $P_{min}$ be the minimal value of $P$ over $R$.  Given
$P_{min}$ we could find the largest $s$
such that $Prob$({number of received single-photon pulses} 
$\leq s| P_{min}) \leq \epsilon.$ The final result is
that in the set of all Bob's detections, at least $s$ detections
came from single-photon pulses with confidence $(1-\epsilon)^{2M +
1}$. However finding the global minimum of a nonlinear function
like $P$ over a complicated region like $R$ is difficult.  
Instead we use a more conservative value $P^\prime _{min} \leq
P_{min}$ to calculate $s$.  We use the lower bounds for $\mu$ and
$y_1$ for the numerator of $P$ and we use upper bounds for $\mu$
and all $y_n$ in the denominator of $P$.  Plugging all these
values into $P$ yields $P^\prime_{min}$ and we then use
$P^\prime_{min}$ to calculate $s$, the bound on single-photon
detections with confidence $(1-\epsilon)^{2M+1}.$

In the process of incorporating this analysis in a full protocol \cite{lanl_protocol}, we would also need to determine the bit error rate for the sifted single photons.  Let $b_n$ be the BER for an $n$-photon pulse prepared by Alice.  If every laser pulse is well-defined in polarization, then we could bound the single-photon BER $b_1$ by calculating upper and lower bounds (with confidence level $1 - \epsilon$) of the observed BER $B_j = e^{-\mu_j} \sum_n \frac{(\mu_j)^n}{n!} b_n y_n$ for each signal strength $\mu_j$ and solve for the largest possible value of $b_1$.

\jimhsection{Protocol of possible implementation}
For purposes of illustration, let us consider an example protocol that is well-suited for a free-space QKD scenario.  The decoy state method works by Alice sending signals at various strengths and Bob counting detector clicks.  However, it is possible to obtain good bounds on the transmission rates of single-photon signals versus multi-photon signals with a fixed, known laser strength $\mu$, provided that Alice can fire any number of her lasers simultaneously.  In the following protocol, we consider Alice's setup to contain four identical lasers, each producing a weak coherent pulse with a distinct polarization ({\it e.g.}, vertical, horizontal, diagonal, and anti-diagonal).  An important assumption we make is that the output of $j$ lasers firing simultaneously with intensity $\mu$ (averaged over all $j$-tuples) is indistinguishable from the output of one laser firing with intensity $j \mu$ (averaged over polarizations), because they are described by the same density matrix \cite{peres_quantum}.

Let $\epsilon$ be a user-defined parameter for security.  

Let $N$ be the number of clock cycles during the quantum transmission portion of a QKD session.

During each clock cycle, Alice generates four random bits.  Each bit is assigned to one of the four lasers, and each laser is fired (simultaneously) if its bit value is one.  

Let $N_j$ be the number of signal pulses sent during the session with $j$ lasers firing simultaneously.  We thus expect $N_0 \approx \frac{N}{16}$, $N_1 \approx \frac{N}{4}$, $N_2 \approx \frac{3N}{8}$, $N_3 \approx \frac{N}{4}$, and $N_4 \approx \frac{N}{16}$.

Bob records all positive detection results (meaning one or more detectors click) for the $N$ clock cycles.  He informs Alice (over an authenticated public channel) which signals yielded positive detections, and then Alice tells Bob how many lasers were fired for each detected signal.

Let $C_j$ be the total number of positive detections recorded for the set of signals produced by $j$ lasers firing simultaneously.

Let $y_n$ be the true conditional transmission probability of an $n$-photon pulse ({\it i.e.} the probability that Bob observes a click when Alice prepares an $n$-photon pulse).  Note that $y_0$ is the detector noise rate (background counts plus dark counts).

Let $Y_j$ be the true conditional transmission probability of a pulse with strength $\mu_j$ ({\it i.e.} the probability that Bob observes a click when Alice fires $j$ lasers simultaneously).  Then $Y_j = e^{-\mu_j} \sum_{n=0}^{\infty} \frac{(\mu_j)^n}{n!} y_n$.  Lo {\it et al} \cite{lo_decoy} use $\frac{C_j}{N_j}$ as a maximum likelihood estimator for $Y_j$, but we will instead consider confidence levels from finite sample statistics. 

Let ${Y}_j^+$ and ${Y}_j^-$ be upper and lower bounds on $Y_j$ at confidence level $1 - \epsilon$, given that Bob observes $C_j$ detections for $N_j$ signals.  The values of ${Y}_j^\pm$ are calculated by solving $\binom{N_j}{C_j} ({Y}_j^\pm)^{C_j} (1-{Y}_j^\pm)^{N_j - C_j} \leq \epsilon$. 

Let $K$ to be the number of variables we will constrain.  For this example, we found $K = 11$ to be sufficient.

Now solve for the minimum value of $y_1$ subject to $0 \leq y_k \leq 1$ and the following set of inequalities:
\begin{eqnarray}
{Y}_j^+
&\geq& e^{-\mu_j} \sum_{k=0}^{K} y_k \frac{(\mu_j)^k}{k!} \\
(1-{Y}_j^-)
&\geq& e^{-\mu_j} \sum_{k=0}^{K} (1-y_k) \frac{(\mu_j)^k}{k!}
\end{eqnarray}
These inequalities are a set of hyperplanes which define the faces of a convex polytope.
 
\begin{figure}[t]
\includegraphics[width=86mm]{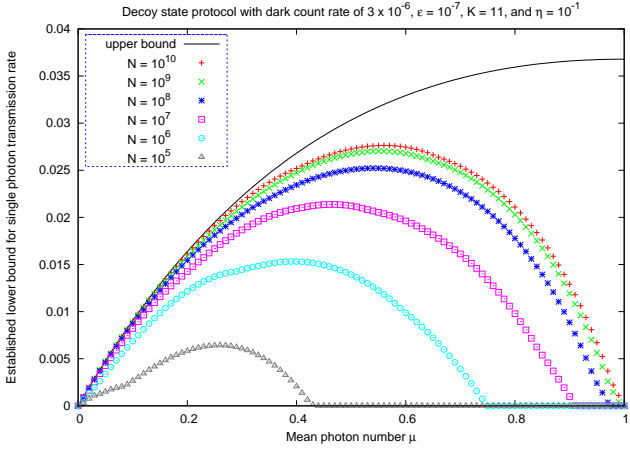}
\caption{\label{fig:4lasers1} (Color online)
\bf Rate of received single-photon signals versus mean photon number $\mu$ over a channel acting as a beamsplitter with transmission $\eta = 10^{-1}$}
\end{figure}

\begin{figure}[t]
\includegraphics[width=86mm]{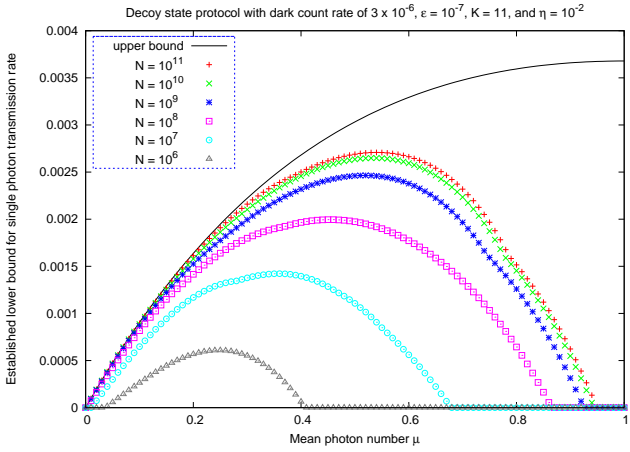}
\caption{\label{fig:4lasers2} (Color online)
\bf Rate of received single-photon signals versus mean photon number $\mu$ over a channel acting as a beamsplitter with transmission $\eta = 10^{-2}$}
\end{figure}

Finally, to determine the single-photon bit error rate, let $b_n$ be the true BER of an $n$-photon pulse.  For this protocol, only the $N_1$ signals prepared by Alice by firing exactly one laser have definite polarization.  Therefore, we only have one signal strength $\mu_1 = \mu$ that can be used to measure the BER in this setup, so a conservative approach would be to let $b_n = 0$ for $n \geq 2$.  This leads to the constraint that $B_1^- \leq e^{-\mu} \left( \frac{1}{2} y_0 + \mu b_1 y_1 \right) \leq B_1^+$,
where $B_1^+$ and $B_1^-$ are upper and lower bounds with confidence $1 - \epsilon$ on the observed BER of the $N_1$ signals.

\begin{figure}[t]
\includegraphics[width=86mm]{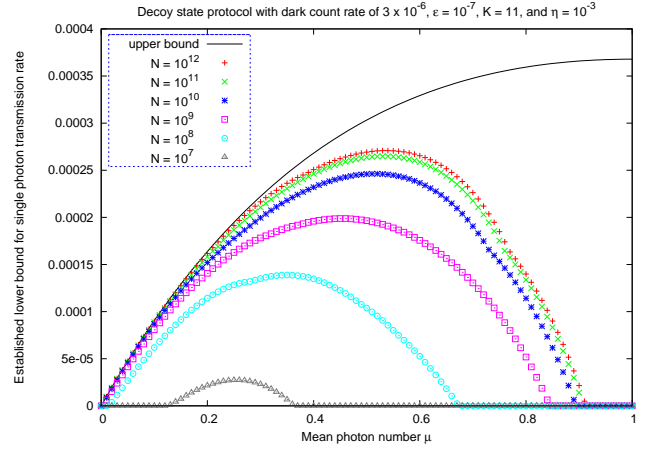}
\caption{\label{fig:4lasers3} (Color online)
\bf Rate of received single-photon signals versus mean photon number $\mu$ over a channel acting as a beamsplitter with transmission $\eta = 10^{-3}$}
\end{figure}

\begin{figure}[t]
\includegraphics[width=86mm]{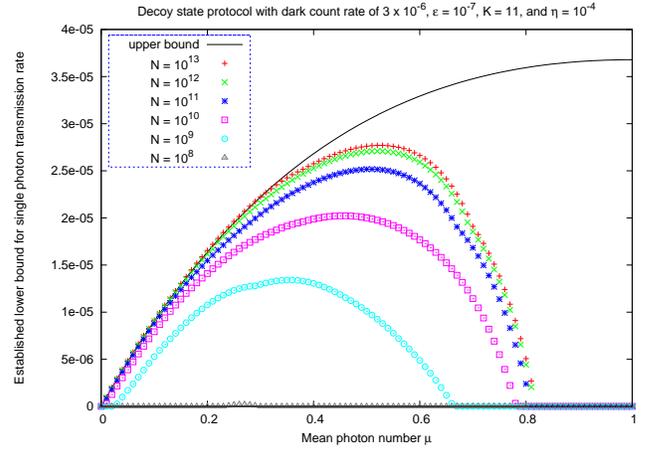}
\caption{\label{fig:4lasers4} (Color online)
\bf Rate of received single-photon signals versus mean photon number $\mu$ over a channel acting as a beamsplitter with transmission $\eta = 10^{-4}$}
\end{figure}

\jimhsection{Numerical results}
Figures \ref{fig:4lasers1}, \ref{fig:4lasers2}, \ref{fig:4lasers3}, and \ref{fig:4lasers4} plot the results of simulations of the protocol just described, with observed transmission efficiency $\eta$ ranging from $10^{-1}$ down to $10^{-4}$.  We chose security parameter $\epsilon = 10^{-7}$ and detector dark count (plus background count) rate $y_0 = 3 \times 10^{-6}$ per clock cycle.  This value of $y_0$ is comparable to the observed rate at nighttime for the 10-km free-space experiment with clock rate of 1 MHz \cite{lanl_10km}.

The optimal mean photon numbers are found to be around 0.35, 0.45, and 0.52 for session size $N = 10^5/\eta$, $N = 10^6/\eta$, and $N = 10^7/\eta$, respectively.  Asymptotically, this protocol has optimal $\mu \sim 0.55$, which can be compared to the asymptotic result of $\mu \sim 0.5$ calculated by Lo {\it et al} \cite{lo_decoy} with similar parameters.  

We require $N \eta \gtrsim 10^5$ to have sufficient statistics for our confidence level of $1 - 10^{-7}$.  However, reducing $\epsilon$ to, say, $10^{-14}$ has minor effects; the lower bounds of the single-photon rates are decreased for $N = 10^5 / \eta$ by less than $25\%$ and for $N = 10^6 / \eta$ by less than $10\%$.

Increasing the dark count (plus background count) rate by a factor of ten has negligible effects on the resulting lower bounds for the single-photon rate.  
We also examined the impact of BER on the resulting secret key bit rate under this protocol.  We found that, roughly speaking, a BER of 7\% assuming optimal individual attacks \cite{opt_indiv} or a BER of 3\% allowing general coherent attacks \cite{shor_preskill} both resulted in halving the secret bit rate and shifting the optimal $\mu$ downwards by about a quarter.  Most of this shift is due to the conservative estimation of the single-photon BER $b_1$ by setting $b_n = 0$ for $n \geq 2$.

\begin{figure}[t]
\includegraphics[width=86mm]{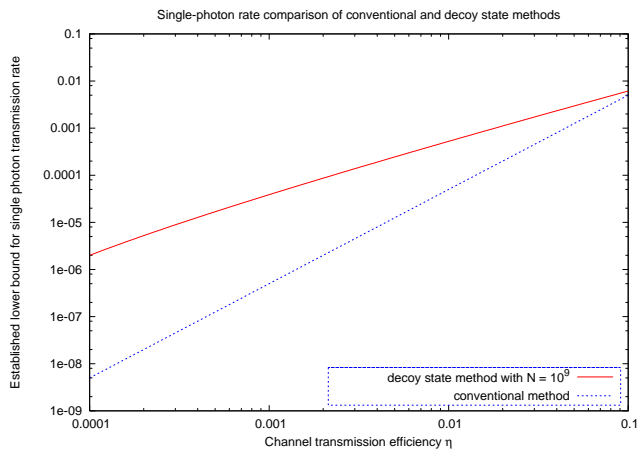}
\caption{\label{fig:RateComparison} (Color online)
\bf Rate comparison before error correction and privacy amplification of decoy state method (solid red line) versus conventional method (dashed blue line) of handling PNS attack.}
\end{figure}

\jimhsection{Rate comparison}
Conventionally \cite{indiv_attacks}, the PNS attack is handled by choosing an appropriately small value for mean photon number $\mu$, so that even if all multi-photon pulses are transmitted perfectly ($y_n = 1$ for ${n \geq 2}$), some single-photon pulses must still remain in the set of Bob's detections.  Then, the guaranteed single-photon rate is close to $R = \mu (\eta - \mu/2)$, which is maximized when we choose $\mu = \eta$.  In Fig. \ref{fig:RateComparison}, the dashed blue line corresponds to this rate as a function of $\eta$.

Suppose we implement the decoy state protocol for a session size of $N = 10^9$ pulses.  The single-photon rate is given by $R = (1/4) f \mu e^{-\mu} \eta$, where $f$ is the fraction of the lower bound to the upper bound in Figs. \ref{fig:4lasers1}--\ref{fig:4lasers4}, and the optimal value of $\mu$ ranges from around 0.55 for $\eta = 0.1$ down to about 0.35 for $\eta = 0.0001$.  In Fig. \ref{fig:RateComparison}, the solid red line corresponds to the rate with these values.  

\jimhsection{Conclusions}
The decoy state method can be implemented with current technology, and it greatly enhances the practical security of quantum key distribution.  In particular, photon number splitting attacks, where Eve has active control of the quantum channel, can be thwarted without drastically reducing the secret bit rate by preparing pulses at various intensities (such as by firing a variable number of lasers with fixed intensity).  We have shown how to incorporate confidence levels from finite statistics into the decoy state method.  Choosing the best distribution and intensities for the set of decoy and signal states is a huge optimization problem, which depends on such values as channel loss, dark count and background count rates, and acceptable security parameters.  However, we have demonstrated that even with a few easily constructed decoy states, high rates of secure QKD can be established with high confidence.

\jimhsection{Acknowledgments}
We gratefully acknowledge helpful discussions with Hoi-Kwong Lo.  We made use of The Geometry Center's Qhull program \cite{qhull} to compute halfspace intersections.  This work was supported by ARDA.
\bibliography{PRL_Decoy}

\end{document}